\documentclass[11pt,a4paper]{JHEP3}
\usepackage{amsfonts,ulem,bbold}   
\usepackage{amssymb,amsmath}%,graphicx,wrapfig,epsfig} 

\newcommand{\be}{\begin{equation}}  
\newcommand{\ee}{\end{equation}}  
\newcommand{\bea}{\begin{eqnarray}}  
\newcommand{\eea}{\end{eqnarray}}  
\newcommand{\nn}{\nonumber}  
\newcommand{\p}{\partial}
\newcommand{\C}{{\cal C}_{(2)}}

\newcommand{\mX}{\mathbb X}

\newcommand{\tf}{{}^{{}^{{}_{3}}}\!\!f}
\newcommand{\tR}{{}^{{}^{{}_{3}}}\!\!R}
\newcommand{\pb}[1]{\big\{{#1}\big\} }
\newcommand{\ds}{\displaystyle}
\DeclareMathOperator{\tr}{tr}  

\title{Confirmation of the Secondary Constraint and Absence of Ghost
  in Massive Gravity and Bimetric Gravity}   

\author{S. F. Hassan\\ Department of Physics \& The Oskar  
Klein Centre,\\ Stockholm University, AlbaNova University Centre,  
SE-106 91 Stockholm, Sweden \\ E-mail: \email{fawad@fysik.su.se}
}  
  
\author{Rachel A. Rosen\\ Physics Department and Institute for
  Strings, Cosmology, and Astroparticle Physics,\\ 
Columbia University, New York, NY 10027, USA \\ E-mail:
\email{rar2172@columbia.edu}}  
  
\abstract{In massive gravity and in bimetric theories of gravity, two
  constraints are needed to eliminate the two phase-space degrees of
  freedom of the Boulware-Deser ghost.  For recently proposed
  non-linear theories, a Hamiltonian constraint has been shown to
  exist and an associated secondary constraint was argued to arise as
  well.  In this paper we explicitly demonstrate the existence of the
  secondary constraint.  Thus the Boulware-Deser ghost is completely
  absent from these non-linear massive gravity theories and from the
  corresponding bimetric theories.}


\preprint{}

\begin{document}

\section{Introduction and summary}

Generic theories of massive gravity and bimetric gravity contain a
Boulware-Deser ghost mode \cite{BD} which renders these theories
unstable. To avoid this inconsistency, the equations of motion of a
healthy, ghost-free massive gravity or bimetric theory must contain
two constraints which eliminate both the ghost field and the momentum
canonically conjugate to it. In \cite{HR3,HR4,HR5} it was established
that one such constraint, the Hamiltonian constraint, exists in a
special class of massive gravity and bimetric theories.   In
\cite{HR3,HR4,HR5} an associated secondary constraint was argued to
arise as well. Here we rigorously prove the existence of the secondary
constraint and obtain an explicit expression for it.

The class of theories being considered are extensions of the massive gravity actions
originally proposed in \cite{dRG, dRGT} where they were formulated
with a flat reference metric. In these works, the proposed actions
were shown to be ghost-free in the decoupling limit and to possess a
Hamiltonian constraint up to fourth order in perturbation theory.  For 
complementary analyses, see \cite{dRGT2,dRGT3}.  For other recent, related work, see \cite{Deffayet:2011uk,Chamseddine:2011mu,Deffayet:2011rh,deRham:2011pt,Folkerts:2011ev,D'Amico:2011jj,Gumrukcuoglu:2011ew,Koyama:2011wx,Alberte:2011ah,Kluson:2011aq}.  For a recent review on massive gravity, see \cite{Kurt}.

The full non-linear analysis of the Hamiltonian constraint was performed in
\cite{HR3,HR4} and is based on a reformulation of these models presented
in \cite{HR}, where they are also extended to an arbitrary reference metric. In
\cite{HR5} a ghost-free bimetric theory based on these massive gravity
theories was proposed in which the reference metric is promoted to a
dynamical field. The existence of the Hamiltonian constraint for the
bimetric theory was also established in \cite{HR5}.  For recent work related to the
bimetric theory, see \cite{Comelli:2011wq,Volkov:2011an,vonStrauss:2011mq}.

More specifically, the analyses of \cite{HR3,HR4,HR5} showed that both
the massive gravity and bimetric theories contain a constraint, say
${\cal C}=0$, that can be regarded as a generalization of the
Hamiltonian constraint in General Relativity. A condition for
consistency is the preservation of this constraint in time, $d{\cal
  C}/dt=0$. This condition will lead to a secondary constraint,
provided the Poisson bracket of $\cal C$ with itself vanishes on the
constraint surface, $\pb{{\cal C}(x),{\cal C}(y)}\approx 0$. In
\cite{HR3,HR4,HR5} this bracket was not computed, but was assumed to
vanish considering that it vanishes in two very different limits of
massive gravity: the zero mass limit, which is General Relativity, and
the linearized limit, which is the linear massive Fierz-Pauli theory
\cite{FP1,FP2}. It was then argued that a non-trivial secondary
constraint should exist in the non-linear theory, as an
extension of the secondary constraint in the Fierz-Pauli theory. This
led to the conclusion that the Boulware-Deser ghost is eliminated by
the two constraints in the non-linear theory.

However, it was recently argued in \cite{Kluson} that in massive
gravity, the Poisson bracket $\pb{{\cal C}(x),{\cal C}(y)}$ does not
vanish at the non-linear level, hence the theory does not contain a
secondary constraint. If true, this would imply that, even though the
ghost field is eliminated by the Hamiltonian constraint, the momentum
canonically conjugate to the ghost is not eliminated. The theory would
thus have an odd dimensional phase space -- an odd situation indeed.
In this paper, we explicitly compute the Poisson bracket in massive
gravity and show that $\pb{{\cal C}(x),{\cal C}(y)}\approx 0$ indeed
holds. Then we explicitly compute the secondary constraint $\C$. We
also argue that no other associated constraints exist. This result is
then generalized from massive gravity to bimetric gravity. This
conclusively proves the complete absence of the Boulware-Deser ghost
both in non-linear massive gravity, as well as in non-linear bimetric 
gravity.

The apparent contradiction of this result with the work \cite{Kluson}
can be understood in the following way. The constraint analysis of
\cite{Kluson} is based on a straightforward counting of Lagrange
multipliers and equations of motion. However, as the existence of the
Hamiltonian constraint itself demonstrates, such a counting can be
misleading. If the equations of motion are degenerate, then an
equation which would otherwise determine a Lagrange multiplier instead
becomes a constraint on the dynamical variables. The analysis of
\cite{Kluson} is insensitive to this possibility, while here we show
that this is indeed the case.

The paper is organized as follows: In section 2 we review non-linear
massive gravity and its Hamiltonian constraint. Section 3 proves the
existence of the secondary constraint and obtains an explicit
expression for it. The absence of higher constraints is argued here as
well. Section 4 extends the results to bimetric gravity.

\section{Review of the Hamiltonian constraint in massive gravity}

In this section we review the ADM formulation of massive gravity and
the derivation of the Hamiltonian constraint based on \cite{HR3,HR4}.
 
\subsection{Massive gravity in ADM variables}

We start by reviewing the derivation of the Hamiltonian constraint of
the massive gravity theories proposed in \cite{dRG,dRGT}, as was
carried out in \cite{HR3,HR4}.  Here, we construct these theories with
respect to an arbitrary but non-dynamical reference metric
$f_{\mu\nu}$.  In the last section of this paper we will discuss the
generalization of these arguments to a theory with a dynamical
$f_{\mu\nu}$.

The most general massive gravity actions can be written as \cite{HR}
\be
\label{act} 
S=M_p^2\int d^4x\sqrt{- \det g}\,\bigg[R +2m^2 \sum_{n=0}^{3} \beta_n\,
  e_n(\sqrt{g^{-1} f})\bigg] , 
\ee 
where the square root of the matrix is defined such that
$\sqrt{g^{-1}f} \sqrt{g^{-1}f} = g^{\mu \lambda}f_{\lambda \nu}$.  The
$e_k(\sqrt{g^{-1} f})$ are elementary symmetric polynomials of the
eigenvalues $\lambda_n$ of the matrix $\sqrt{g^{-1} f}$.  For a $4
\times 4$ matrix $\mX$ and using the notation $[M]=\tr M$, they can be
written as, 
 \bea
\label{e}   
e_0(\mX)&=& 1  \, , \nonumber \\  
e_1(\mX)&=& [\mX]  \, ,\nonumber \\  
e_2(\mX)&=& \tfrac{1}{2}([\mX]^2-[\mX^2]), 
\label{e_n}  \\  
e_3(\mX)&=& \tfrac{1}{6}([\mX]^3-3[\mX][\mX^2]+2[\mX^3])
\, ,\nonumber \\   
e_4(\mX)&=&\tfrac{1}{24}([\mX]^4-6[\mX]^2[\mX^2]+3[\mX^2]^2   
+8[\mX][\mX^3]-6[\mX^4])\, ,\nonumber \\  
e_k(\mX)&=& 0 \qquad {\rm for} \quad k>4 \, . \nonumber
\eea 
The $\beta_n$ correspond to four free parameters, two of which are the
graviton mass and the cosmological constant. 

The physical content of these theories is most easily identified in
the ADM formulation \cite{ADM} which is based on a 3 + 1 decomposition
of the metric.  Let $N$ and $N_i$ denote the lapse and shift functions
of the metric $g_{\mu\nu}$ while $L$ and $L_i$ denote the lapse and
shift functions of the metric $f_{\mu\nu}$, so that
\begin{align} 
g_{\mu\nu}:\qquad & N =(-g^{00})^{-1/2}\,, \qquad N_i = g_{0i}\,, 
\qquad \gamma_{ij}=g_{ij}\,, \label{gADM}\\[.1cm]
f_{\mu\nu}:\qquad & L =(-f^{00})^{-1/2}\,, \qquad L_i = f_{0i}\,, 
\qquad \tf_{ij}=f_{ij}\,\label{fADM}.  
\end{align}
In massive gravity $f_{\mu\nu}$ is non-dynamical. The lapse and
shift variables $N_\mu$ appear without time derivatives in 
(\ref{act}) and are thus non-dynamical as well. The components
$\gamma_{ij}$ are dynamical and, along with the their canonically
conjugate momenta $\pi^{ij}$, constitute $12$ phase-space degrees of
freedom, i.e., six potentially propagating modes. Of these, five
correspond to the massive graviton while the sixth one, if not
eliminated by the constraint equations, is the
Boulware-Deser ghost.

In terms of these variables, the Lagrangian in (\ref{act}) becomes, 
\be
\label{L}
{\cal L} = \pi^{ij}\partial_t \gamma_{ij} + NR^0+R_i N^i
+2m^2\sqrt{\det\gamma}\,N\,\sum_{n=0}^{3}\beta_n\,e_n(\sqrt{g^{-1}f})\,,
\ee
where, as in General Relativity, 
\be
R^0= \sqrt{\det \gamma}\,\,\tR+\frac{1}{\sqrt{\det\gamma}}
\left(\tfrac{1}{2}\pi^i_{~i}\pi^j_{~j}-\pi^{ij}\pi_{ij}\right)\,,
\qquad
R_i = 2\sqrt{\det\gamma}\,\gamma_{ij} \nabla_k(\frac{\pi^{jk}}
{\sqrt{\det\gamma}}) \,. 
\label{Rmu}
\ee
The $N_\mu$ equations of motion derived from (\ref{L}) contain a
single constraint on the variables $\gamma_{ij}$ and $\pi^{ij}$. This
is the Hamiltonian constraint which is necessary (though not 
sufficient) to remove the ghost. However, since the potential in
(\ref{L}) is highly nonlinear in the lapse $N$, the existence of a
Hamiltonian constraint is far from apparent. To make this constraint
manifest, we have to work with the appropriate set of variables.  We
define these in the following subsection. 

\subsection{The new shift-like variables}
The existence of the Hamiltonian constraint is due to the fact that
the four $N_\mu$ equations of motion in fact depend on only three 
functions $n^i$ (for $i=1,2,3$) of the four variables $N_\mu$
\cite{HR3,HR4}.  Thus three combinations of these equations determine 
the $n^i$ and the remaining equation is the Hamiltonian
constraint. This can be made explicit if we work with the new
shift-like variables $n^i$ which are related to the ADM variables of
(\ref{gADM}) and (\ref{fADM}) through (see \cite{HR4} for details), 
\be
N^i - L^i  =\left(L\,\delta^i_{~j} + N\, D^i_{~j}\right) n^j\,,
\label{NnD}
\ee
where $N^i=\gamma^{ij}N_j$ and $L^i=\tf^{ij}L_j$. The matrix $D =
D^i_{\,j}$ is determined by the condition,
\be
\sqrt{x}\,D =\sqrt{(\gamma^{-1}-D n n^T D^T)\,\tf} \, ,
\label{D}
\ee 
where,
\be
x \equiv 1 -n^i\,(\tf_{ij})\,n^j\,.
\label{x}
\ee
In (\ref{D}) we have introduced the notation $n$ for the column vector
$n^i$, and $n^T$ for its transpose.  We will occasionally use this
matrix notation when the meaning is clear.  The matrix $D$ has the
following important property
\be
\label{fD}
\tf_{ik}D^k_{~j} = \tf_{jk}D^k_{~i} \, .
\ee
This identity will be used often to simplify our
calculations. Although $D$ can be determined in terms of $n^i$, the
explicit solution is not needed here.

We can use (\ref{NnD}) to eliminate the shifts $N^i$ in favor of
$n^i$. The resulting form of the Lagrangian (\ref{L}) is then linear
in $N$.  It can be compactly written as,
\be
{\cal L} = \pi^{ij}\partial_t \gamma_{ij} - {\cal H}_0+N {\cal C} \, .
\label{LADM}
\ee
Here ${\cal H}_0$ and ${\cal C}$ stand for 
\begin{align}
{\cal H}_0 &= -(L\,n^i+L^i)R_i-2m^2\,L\,\sqrt{\det \gamma} \, U, 
\label{H0}\\[.1cm]
{\cal C} &= R^0+R_i D^i_{\,j}n^j+2m^2\sqrt{\det \gamma} \, V\,.
\label{C}
\end{align}
The functions $U$ and $V$ appearing above are given by the following 
expressions, 
\begin{align}
U & \equiv  \beta_1 \sqrt{x} +\beta_2\left[\sqrt{x}^{\,2}\,D^i_{~i}
+n^i \,\tf_{ij}D^j_{~k}n^k\right]  \nn\\
&\quad +\beta_3\left[\sqrt{x}\, (D^l_{~l}\,n^i \, \tf_{ij}D^j_{~k}n^k
-D^i_{~k}\,n^k \, \tf_{ij}D^j_{~l}n^l)+\tfrac{1}{2}\sqrt{x}^{\,3} 
(D^i_{~i}D^j_{~j}-D^i_{~j}D^j_{~i}) \right]\,, \label{U} \\[.1cm]
V &\equiv \beta_0+\beta_1 \sqrt{x} \,D^i_{~i}+\tfrac{1}{2}\beta_2
\sqrt{x}^{\,2}\,\left[D^i_{~i}D^j_{~j}-D^i_{~j}D^j_{~i})\right]\nn\\
&\hspace{3.7cm}+\tfrac{1}{6}\beta_3\sqrt{x}^{\,3}\,\left[
D^i_{~i}D^j_{~j}D^k_{~k}-3D^i_{~i}D^j_{~k}D^k_{~j}+2D^i_{~j}D^j_{~k}D^k_{~i}
\right] \, .\label{V}
\end{align}
The massive gravity Lagrangian is now given entirely in terms of
$\gamma_{ij}, \pi^{ij}, N$ and $n^i$. It also involves the ADM
parameters of $f_{\mu\nu}$ (\ref{fADM}) as non-dynamical fields.

\subsection{The Hamiltonian constraint}

In the Lagrangian (\ref{LADM}), the four variables $N$ and $n^i$ are
non-propagating while the six modes described by ($\gamma_{ij},
\pi^{ij}$) potentially propagate and include the ghost field and its
conjugate momentum. The Hamiltonian constraint is identified after
imposing the $n^k$ equations of motion. These were derived in
\cite{HR3,HR4}.  Here we outline the derivation. Varying $\cal L$ in
(\ref{LADM}) with respect to $n^k$ gives the equations of motion, 
\be
\label{Ldn}
\frac{\p\cal L}{\p n^k}=-\frac{\p{\cal H}_0}{\p n^k}
+N \frac{\p\cal C}{\p n^k}=0\,.
\ee
To determine the variation of the two terms in (\ref{Ldn}), it is
helpful to use the following relations, derived from the expression
(\ref{D}), 
\begin{align}
&\frac{\p}{\p n^k}\tr(\sqrt{x}\,D) = - \frac{1}{\sqrt{x}}\,n^T\,\tf\, 
\frac{\p(Dn)}{\p n^k}\, , \nn \\[.1cm] 
&\frac{\p}{\p n^k}\tr(\sqrt{x} \,D) ^2 = -2\,n^T\,\tf\,D\,
\frac{\p(Dn)}{\p n^k}\,,\label{dBs} \\[.1cm] 
&\frac{\p}{\p n^k}\tr(\sqrt{x} \,D)^3 = -3\sqrt{x}\,n^T\,\tf\,D^2\,
\frac{\p(Dn)} {\p n^k}\, .\nn
\end{align}
One then obtains the useful expressions,
 \be
\frac{\p{\cal H}_0}{\p n^k}=-L\, {\cal C}_k\,,\qquad\qquad\qquad
\frac{\p\cal C}{\p n^k}={\cal C}_i\, \frac{\p (D^i_{~j}n^j)}{\p n^k} \, .
\label{HCn}
\ee
Note that both these variations are proportional to ${\cal C}_i$, given
by, 
\begin{align}
{\cal C}_i & = R_i-2m^2\sqrt{\det\gamma}\,\frac{n^lf_{lj}}{\sqrt{x}}
\bigg[\beta_1\,\delta^j_{i} +\beta_2\,\sqrt{x}\, (\delta^j_{i} 
D^m_{~m}-D^j_{~i})\nn \\
&\hspace{2.15cm}+\beta_3 \, \sqrt{x}^{\,2} \, \left(\tfrac{1}{2}\delta^j_{~i} 
(D^m_{~m} D^n_{~n}\, -\,D^m_{~n} D^n_{~m}) +D^j_{~m} D^m_{~i}\, -\,
D^j_{~i} D^m_{~m}\right) \bigg] \, .
\label{Ci}
\end{align}
Thus the variation of ${\cal L}$ becomes,
\be
\frac{\p\cal L}{\p n^k}={\cal C}_i\, \Big[L\,\delta^i_k + N \frac{\p
    (D^i_{~j}n^j)}{\p n^k} \Big]=0 \, .
\label{CiSB}
\ee
The matrix within the square brackets is the Jacobian of the
 transformation (\ref{NnD}) and is invertible.  Hence the $n^i$
equations of motion are 
\be
{\cal C}_i(\gamma,\pi,n)=0 \, .
\label{neom}
\ee
These are independent of the lapse $N$ and can in principle be solved
to determine $n^i$ in terms of $\gamma_{ij}$ and $\pi^{ij}$. So far,
the explicit solution is known only for the minimal massive action
corresponding to $\beta_2=\beta_3=0$. But, as shown below, the
equations (\ref{neom}) alone are enough to demonstrate the absence of
the ghost. Note that (\ref{neom}) implies the vanishing of the
individual variations in (\ref{HCn}). 

Now, varying the Lagrangian with respect to $N$ gives, 
\be
{\cal C}(\gamma,\pi,n) = 0 \, .
\label{NeomC}
\ee
On eliminating the $n^i$ using the solutions\footnote{That this is
  always possible follows from the fact that when the $n^i$ equations
  of motion are satisfied we have $\p C/\p n^i=0$. Hence on the space
  of solutions, $\cal C$ depends only on $\gamma_{ij}$ and
  $\pi^{ij}$.} of (\ref{neom}), this equation becomes a constraint on
the 12 dynamical variables $\gamma_{ij}$ and $\pi^{ij}$, reducing the
number of independent degrees of freedom to 11. This is the
Hamiltonian constraint.  The degree of freedom it eliminates is the
ghost field.  One more constraint is necessary to eliminate the
momentum canonically conjugate to the ghost field. This is obtained in  
the next section.

\section{The secondary constraint}

In this section we show that the Hamiltonian constraint gives rise to
a secondary constraint and we obtain its explicit expression. This new
constraint eliminates the momentum canonically conjugate to the ghost.
Thus there remain 10 degrees of freedom out of an original 12.

\subsection{Existence of a secondary constraint}

In order to be consistent, the Hamiltonian constraint ${\cal C}=0$ must
be preserved under the time evolution of $\gamma_{ij}$ and $\pi^{ij}$,
that is, $d{\cal C}/dt=0$. This requirement can be implemented in the 
Hamiltonian formulation. From the Lagrangian (\ref{LADM}), the
Hamiltonian is given by,  
\be
H = \int d^3 x \, ({\cal H}_0-N {\cal C}) \,.
\label{H}
\ee
In this expression the $n^i$ are determined by their equations of
motion (\ref{neom}), while $N$ is a Lagrange multiplier.  The only
dynamical variables are $\gamma_{ij}$ and $\pi^{ij}$. 

In the Hamiltonian formulation, the consistency condition on the time
evolution of the constraint becomes,  
\be
\frac{d}{dt}\, {\cal C}(x) = \pb{{\cal C}(x),H} = 0 \,,
\label{ddtC}
\ee
where the Poisson bracket is defined as,
\be
\pb{{\cal C}(x),H} =\int d^3z \left(\frac{\delta\,{\cal C}(x)}
{\delta\gamma_{mn}(z)}\,\frac{\delta H}{\delta\pi^{mn}(z)}
-\frac{\delta\,{\cal C}(x)}{\delta\pi^{mn}(z)}\,
\frac{\delta H}{\delta\gamma_{mn}(z)}\right) \, .
\label{pb-def}
\ee
The $\delta$ denote total variations, including dependence on
$\gamma_{ij}$ and $\pi^{ij}$ through the $n^k$. Using (\ref{H}), the
consistency condition (\ref{ddtC}) becomes, 
\be
\pb{{\cal C}(x),H}=\int d^3 y \, \pb{{\cal C}(x),{\cal H}_0(y)}-
 \int d^3 y \,N(y) \, \pb{{\cal C}(x),{\cal C}(y)} =0 \, .
\label{PBCH}
\ee
It is necessary for this condition to hold only on the constraint
surface, i.e., when ${\cal C} = 0$. 

When imposing the above consistency condition, two possibilities could
arise:  
\begin{enumerate}
\item $\pb{{\cal C}(x),{\cal C}(y)}\not\approx 0$, where the symbol
  $\approx$ is used for equalities that hold on the constraint
  surface, i.e., when ${\cal C}=0$. In this case (\ref{PBCH}) becomes
  an equation for $N$ and does not act as a constraint on
  $\gamma_{ij}$ and $\pi^{ij}$. This leads to a theory with an odd
  dimensional phase space, or equivalently, with $5.5$ propagating
  modes.  Such a possibility was raised in \cite{Kluson} and was
  argued to be the case in massive gravity.
\item $\pb{{\cal C}(x),{\cal C}(y)}\approx 0$.  In this case, the
  condition (\ref{PBCH}) is independent of $N$ and thus acts as a
  secondary constraint on $\gamma_{ij}$ and $\pi^{ij}$, 
\be
\label{C2}
\C(x)= \pb{{\cal C}(x),H_0}=0\,.
\ee
where, $H_0=\int d^3 y\,{\cal H}_0(y)$. The secondary constraint
eliminates the momentum canonically conjugate to the ghost field and
leads to a theory with 5 propagating modes.  The Boulware-Deser ghost 
mode is thus entirely absent.  This is the scenario advocated in
\cite{HR3,HR4,HR5}. 
\end{enumerate}
In this section we will explicitly show that massive gravity indeed
corresponds to the second case given above and is, therefore, entirely
free of the Boulware-Deser ghost.  We will show in section four that
this argument also holds for the theories of bimetric gravity based on
these massive gravity actions \cite{HR5}. 

To prove the existence of the secondary constraint, we now show that, 
\be
\label{pbCC}
\pb{{\cal C}(x),{\cal C}(y)}\approx 0 \, .
\ee
To evaluate the Poisson bracket, note that ${\cal C}$ depends on
$\gamma_{ij}$ and $\pi^{ij}$ explicitly as well as implicitly, through
the $n^i$. However, from (\ref{HCn}) it follows that when the $n^i$
equations of motion are satisfied one has $\tfrac{\delta\,{\cal
    C}}{\delta n^i}=0$. Thus the most general variation with respect
to the canonical variables is given by,
\be
\delta\, {\cal C}=\frac{\delta\,{\cal C}}{\delta\gamma_{mn}}\bigg\vert_{n^i}\, 
\delta\gamma_{mn}+\frac{\delta\,{\cal C}}{\delta\pi^{mn}}\bigg\vert_{n^i}\,
\delta\pi^{mn}\,,
\label{varCH}
\ee
where the derivatives are now evaluated at fixed $n^i$. This means
that the Poisson bracket (\ref{pbCC}) can be evaluated by
replacing the total variations $\delta$ by the partial ones evaluated
at fixed $n^i$. From now on we only consider these types of
variations.  

Now we can compute $\pb{{\cal C}(x),{\cal C}(y)}$ using the expression
(\ref{C}) for $\cal C$.  To simplify, note that $R^0$ and $V$ depend,
respectively, only on $\pi$ and $\gamma$ and not on their derivatives.
Thus the term $\pb{R^0(x),\sqrt{\det\gamma}\, V(y)}$ is proportional
to the Dirac delta function $\delta^3(x-y)$.  The net contribution of
this term and its conjugate, $-\pb{R^0(y),\sqrt{\det\gamma}\, V(x)}$,
to the bracket $\pb{{\cal C}(x),{\cal C}(y)}$ is therefore zero, due to the antisymmetry of the bracket.  Taking this into
account gives,
\begin{align}
\pb{C(x), C(y)}&= \pb{R^0(x), R^0(y)} 
+\pb{R_i(x), R_j(y)}\,D^i_{~k}n^k(x) \,D^j_{~l}n^l(y) 
\nn\\[.2cm]
&+\pb{R^0(x),R_i(y)}D^i_{~k}n^k(y)-\pb{R^0(y),R_i(x)}\,D^i_{~k}n^k(x) 
\nn\\[.1cm]
&+ S^{mn}(x)\,\frac{\delta R_i(y)}{\delta\pi^{mn}(x)}\,D^i_{~k}n^k(y)
-S^{mn}(y)\,\frac{\delta R_i(x)}{\delta\pi^{mn}(y)}\,D^i_{~k}n^k(x)\,.
\label{CC-S}
\end{align}
Here, the quantity $S^{mn}$ stands for 
\be
S^{mn}(x)=R_j(x)\frac{\delta (D^j_{~r}n^r)}{\delta\gamma_{mn}}(x)
+2m^2 \frac{\delta(\sqrt{\det\gamma} V)}{\delta\gamma_{mn}}(x)\,.
\label{S}
\ee

The expression (\ref{CC-S}) involves the standard Poisson brackets
of General Relativity (see, for example, \cite{KMT}),
\begin{align}
\pb{R^0(x), R^0(y)}&=-\left[R^i(x)\tfrac{\p}{\p x^i}\delta^3(x-y)
-R^i(y)\tfrac{\p}{\p y^i}\delta^3(x-y)\right]\,,  
\nn \\[.1cm] 
\pb{R^0(x), R_i(y)}&= - R^0(y)\tfrac{\p}{\p x^i}\delta^3(x-y)\,,
\label{pbRR}\\[.1cm] 
\pb{R_i(x), R_j(y)}&= -\left[R_j(x)\tfrac{\p}{\p x^i}\delta^3(x-y)
-R_i(y)\tfrac{\p}{\p y^j}\delta^3(x-y)\right]\,.  \nn
\end{align}
These terms as well as the remaining terms in (\ref{CC-S}) involve
derivatives of the delta function.  In order to manipulate these
unambiguously, we introduce localized smoothing functions $f(x)$ and
$g(y)$, and define  
\be
F \equiv \int d^3x \, f(x)\, {\cal C}(x)\,,\qquad G \equiv \int d^3y
\, g(y)\, {\cal C}(y)\, .  
\ee
Then, the Poisson bracket $\{F,G\}$ is given by 
\be
\{F,G\} = \int\!d^3x\!\int\!d^3y\,f(x)\,g(y) 
\pb{{\cal C}(x),{\cal C}(y)} \, .
\label{FG} 
\ee
We will first compute this bracket and then from it extract $\pb{{\cal
    C}(x),{\cal C}(y)}$. 

To determine the last two terms in (\ref{CC-S}), we note that, on
using the expression (\ref{Rmu}) for $R_j$, for any vector field
$v^j=\gamma^{jk}v_k$ we have,  
\be
\int d^3x \,\frac{\delta R_j(x)}{\delta\pi^{mn}(y)}\, v^j(x) = 
-\left[\nabla_m v_n(y)+\nabla_n v_m(y)\right] \, .
\label{dRdpi}
\ee
Using (\ref{pbRR}) and (\ref{dRdpi}) in expression (\ref{CC-S}), and
then carrying out one of the integrals in (\ref{FG}), one gets, 
\begin{align}
\{F,G\} = -\int d^3x \Big[f \, R^i \,\p_i g &+ f\, \p_i(g \, R^0
  D^i_{~k}n^k) + f\, R_i D^j_{~k}n^k \p_j(g\, D^i_{~l}n^l)  \nn\\
&+2 f \,S^{mn}\nabla_m(g\, \gamma_{nj}D^j_{~k}n^k) 
-(f\leftrightarrow g) \Big] \, ,
\end{align}
where everything under the integral is now a function of $x$.  Again,
due to the antisymmetry of the right-hand-side under the interchange
of $f$ and $g$, the only terms that do not cancel are the ones where 
a derivative directly acts on $f$ or $g$.  These can be written as 
\be
\{F,G\} = -\int d^3x \Big(f \, \p_i g-g\, \p_i f\Big)\, P^i\,, 
\label{CCP}
\ee
where, 
\be
P^i=(R^0 + R_j D^j_{~k}n^k) D^i_{~l}n^l +R^i+2 S^{il}\gamma_{lj}
D^j_{~k}n^k \, .
\label{P1}
\ee
To extract the Poisson bracket $\pb{{\cal C}(x),{\cal C}(y)}$ from
(\ref{CCP}), we write $g(x)=\int d^3y\,g(y)\, \delta^3(x-y)$ and
perform an analogous rewriting of $f$. Then, comparing with (\ref{FG})
gives, 
\be
\pb{{\cal C}(x),{\cal C}(y)}= - \left[
P^i(x)\,\frac{\p}{\p x^i}\delta^3(x-y) 
-P^i(y)\,\frac{\p}{\p y^i}\delta^3(x-y)\right] \, .
\label{CC}
\ee

The expression for $P^i$ can now be simplified. This is the
key step in our proof of the existence of the secondary
constraint. First, in the expression for $S^{mn}$ (\ref{S}), we
eliminate $R_j$ in favor of ${\cal C}_j$ using (\ref{Ci}).  We then
compute $\p V/\p\gamma_{mn}$ using the following identities that are
derived from the relation (\ref{D}),  
\be
\label{dBgs}
\begin{array}{lcl}
\ds
\frac{\p}{\p \gamma_{mn}}\tr(\sqrt{x} \,D) &=&-\ds\frac{1}{\sqrt{x}}
\left(n^T\,\tf\,\frac{\p(Dn)}{\p  \gamma_{mn}}-\frac{1}{2} \tf \,
D^{-1} \,\frac{\p \gamma^{-1}}{\p \gamma_{mn}} \right) \, , 
\\[.5cm]  
\ds\frac{\p}{\p  \gamma_{mn}}\tr(\sqrt{x} \,D)^2&=&-2\ds\left(n^T\,
\tf\,D\,\frac{\p(Dn)}{\p  \gamma_{mn}}-\frac{1}{2} \tf \, \frac{\p
  \gamma^{-1}}{\p \gamma_{mn}}\right) \,,   
\\[.5cm] 
\ds\frac{\p}{\p  \gamma_{mn}}\tr(\sqrt{x} \,D)^3&=&-3\ds\sqrt{x}
\left(n^T\,\tf\,D^2\,\frac{\p(Dn)}{\p  \gamma_{mn}}-\frac{1}{2}  \tf
\, D \, \frac{\p \gamma^{-1}}{\p \gamma_{mn}} \right)\, . 
\end{array}
\ee
This gives,
\be
S^{mn}(x) = 
{\cal C}_i \, \frac{\p (D^i_{\,j}n^j)}{\p \gamma_{mn}} + m^2
\sqrt{\det \gamma} \, \left(V \gamma^{mn} - \bar{V}^{mn} \right) \, , 
\ee
where,
\begin{align}
\bar{V}^{mn}\equiv &\gamma^{mi}\,\Bigg[\beta_1\,\frac{1}{\sqrt{x}}\,
  \tf_{ik} \, (D^{-1})^k_{~j} + \beta_2\,\left(\tf_{ik} \,
  (D^{-1})^k_{~j}\,D^l_{~l}-\tf_{ij}\right) \nn \\
&+\, \beta_3 \, \sqrt{x} \, \left(\tf_{ik} \, D^k_{~j}- \tf_{ij} \,
  D^k_{~k} +\frac{1}{2}\, \tf_{ik} \, (D^{-1})^k_{~j}\, (D^l_{~l}
  \,D^h_{~h}-D^l_{~h}\,D^h_{~l})\right) \Bigg] \,\gamma^{jn}\,.
\label{Vbar}
\end{align}
Since ${\cal C}_i$ is zero by the $n^i$ equations of motion, this is
simply 
\be
S^{mn}(x) = m^2 \sqrt{\det \gamma} \, \left(V \gamma^{mn} -
\bar{V}^{mn} \right) \, . 
\label{S2}
\ee
Thus $P^i$ (\ref{P1}) becomes,
\be
P^i=\left(R^0 + R_j D^j_{~k}n^k+2m^2 \sqrt{\det \gamma} \, V\right)
D^i_{~l}n^l +R^i-2 m^2 \sqrt{\det \gamma} \, \bar{V}^{il}\gamma_{lj}
D^j_{~k}n^k \, . 
\label{P2}
\ee
The expression within the parentheses is simply $\cal C$ (\ref{C}).
Also, from the expression for $\bar{V}^{mn}$ (\ref{Vbar}) and on using
(\ref{fD}), it is easy to verify that the remaining terms give ${\cal
  C}_i$ (\ref{Ci}). Thus, (\ref{P2}) is in fact, $P^i={\cal
  C}\,D^i_{~l}n^l + {\cal C}_l \, \gamma^{li}$. Again, since 
${\cal C}_i$ is zero on the $n^i$ equations of motion, this is
equivalent to  
\be
P^i={\cal C}\,D^i_{~l}n^l \, .
\label{P4}
\ee
Thus the Poisson bracket (\ref{CC}) finally becomes
\be
\pb{{\cal C}(x),{\cal C}(y)}= - \left[
{\cal C}(x)\,D^i_{~j}n^j(x)\,\frac{\p}{\p x^i}\delta^3(x-y)
-{\cal C}(y)\,D^i_{~j}n^j(y)\,\frac{\p}{\p y^i}\delta^3(x-y) \right]\,.
\label{CC2}
\ee
As ${\cal C}(x) = 0$ on the constraint surface, it is clear that, 
\be
\pb{{\cal C}(x),{\cal C}(y)}\approx 0\, .
\ee
This corresponds to case 2 discussed following equation (\ref{PBCH})
and proves the existence of a secondary constraint.  

\subsection{Evaluation of the secondary constraint}
The consistency condition $d{\cal C}/dt=0$ now acts as a secondary
constraint given by (\ref{C2}),  
\be
\C(x)= \int d^3y\pb{{\cal C}(x),{\cal H}_0(y)}=0\,.
\label{C2-y}
\ee
To compute the Poisson bracket, note again equation
(\ref{HCn}),
 \be
\frac{\p{\cal H}_0}{\p n^k}=-L\, {\cal C}_k\,,\qquad\qquad\qquad
\frac{\p\cal C}{\p n^k}={\cal C}_i\, \frac{\p (D^i_{~j}n^j)}{\p n^k} \, .
\ee
Given that ${\cal C}_i$ vanishes on the $n^i$ equations of motion, we
can therefor consider variations of both ${\cal C}$ and ${\cal H}_0$
at fixed $n^i$. Then, using the expressions for ${\cal H}_0$ and $\cal
C$ in (\ref{H0}) and (\ref{C}), we obtain,
\begin{align}
\!\pb{{\cal C}(x) , {\cal H}_0(y)}&
\!=\!-\pb{\!R^0(x),R_i(y)\!}(Ln^i\!+\!L^i)(y)-(D^i_{~k}n^k)(x)\,
\pb{\!R_i(x),R_j(y)\!}(Ln^j\!+\!L^j)(y) \nn\\[.2cm]
&\!+m^2L\,\frac{\delta R^0(x)}{\delta\pi^{mn}(y)}\,\sqrt{\det\gamma}\,U^{mn}(y)
+m^2 L\,(D^i_{~k}n^k)(x)\,\frac{\delta R_i(x)}{\delta\pi^{mn}(y)}
\,\sqrt{\det\gamma}\,U^{mn}(y)\nn\\[.2cm]
&\!-S^{mn}(x)\,\frac{\delta R_i(y)}{\delta\pi^{mn}(x)}\,(Ln^i\!+\!L^i)(y)\,.
\label{CxH0y}
\end{align}
where we have introduced the notation,
\be
U^{mn}\equiv \frac{2}{\sqrt{\det \gamma}}\,
\frac{\delta\left(\sqrt{\det \gamma}\,U\right)}{\delta\gamma_{mn}}\,.
\ee
$S^{mn}$ is given by (\ref{S}) and again reduces to the expression
(\ref{S2}). 

The brackets in the first line of (\ref{CxH0y}) are the brackets of
General Relativity given in (\ref{pbRR}). Under the integral of
(\ref{C2-y}), the ordinary derivatives appearing in (\ref{pbRR}) can
be consistently converted to covariant derivatives. In addition, using
(\ref{Rmu}) and (\ref{dRdpi}), we find
\bea
\frac{\delta R^0(x)}{\delta\pi^{mn}(y)} & = & \frac{1}{\sqrt{\det 
 \gamma}}\left(\gamma_{mn}(x) \pi^k_{~k}(x)-2 \pi_{mn}(x)\right)
\delta^3(x-y)\, ,\\ 
\frac{\delta R_i(x)}{\delta\pi^{mn}(y)}&=& -\big(\gamma_{im}(x)
\nabla_{y^n}+\gamma_{in}(x)\nabla_{y^m} \big) \delta^3(x-y) \, .
\eea
To evaluate the secondary constraint we perform the integration of
(\ref{C2-y}). The derivatives of $\delta^3(x-y)$ that appear in the
above expressions can be manipulated under the integral. After
performing various integrations by parts to move the derivatives from
the delta functions and after dropping the corresponding boundary
terms, we determine the secondary constraint (\ref{C2-y}) to be,
\begin{align}
\C \approx ~&
m^2\, L \left(\gamma_{mn}\,\pi^k_k-2\pi_{mn}\right)\,U^{mn} 
+2m^2\,L\sqrt{\det \gamma}\, (\nabla_m U^{mn})\gamma_{ni} D^i_{~k} n^k 
\nn \\
&+\left(R_j D^i_{~k}n^k -2m^2\sqrt{\det \gamma}\,\gamma_{jk}\bar V^{ki}
\right) \nabla_i(Ln^j+L^j)
\nn\\
&+\sqrt{\det \gamma}\left[\nabla_i\left(\frac{R^0}{\sqrt{\det 
\gamma}}\right) + \nabla_i\left(\frac{R_j}{\sqrt{\det \gamma}}
\right)D^j_{~k}n^k\right](Ln^i+L^i)\,.
\label{C2Exp}
\end{align}
Here we have dropped a term ${\cal C} \, \nabla_i(Ln^i+L^i)$ as this
vanishes on the constraint surface. That the entire secondary
constraint does not somehow vanish on the constraint surface can be
seen by considering the form of the first term in (\ref{C2Exp}) which
appears nowhere in either ${\cal C}$ or the equation of motion ${\cal
  C}_i = 0$.

Now the consistency condition $\C=0$ can be used to eliminate the
component of $\pi^{ij}$ that is canonically conjugate to the ghost
component of $\gamma_{ij}$, which was itself eliminated by the primary
constraint ${\cal C}=0$. This leaves only five propagating modes,
establishing the absence of the Boulware-Deser ghost in massive
gravity.

\subsection{Absence of a tertiary constraint}

It is important to verify that no further constraints are generated by
the secondary constraint and thus no additional degrees of freedom are
eliminated\footnote{There are instances in which we expect massive
  gravity to have fewer than five propagating modes. For instance,
  massive gravity in the background of de Sitter spacetime is known to
  have only four propagating modes when the Higuchi bound is saturated
  \cite{Higuchi,DW}. Such a scenario is describable by the framework
  presented here. However, we expect that in these cases, the
  additional mode is removed by an additional gauge symmetry and not
  by tertiary and quaternary constraints.}. In other words, the
consistency condition
\be
\frac{d}{dt}\, \C(x) = \pb{\C(x),H} = 0 \,,
\label{ddtC2}
\ee
should determine $N$ as a function of the remaining variables, rather
than act as a constraint on the dynamical variables. For this to be
the case, we must have both
\be
\pb{\C(x),{\cal H}_0(y)} \not\approx 0\, ~~{\rm and} ~~
\pb{\C(x),{\cal C}(y)} \not\approx 0 \, .
\ee
For both conditions, it is sufficient to consider the nonlinear theory
at lowest order in $\gamma$ and $\pi$. The theories considered here
reproduce the Fierz-Pauli Hamiltonian at lowest order, by
construction. Both constraints ${\cal C}$ and $\C$ reduce to the
Fierz-Pauli primary and secondary constraints at lowest order. Since
the above two conditions are satisfied by the Fierz-Pauli theory, they
will also be satisfied by the full nonlinear theory. Thus the above
consistency condition (\ref{ddtC2}) becomes an equation for the lapse
$N$, rather than a tertiary constraint.

\section{Extension to bimetric gravity}

It is straightforward to extend the above analysis to the case of
bimetric gravity in which the reference metric $f_{\mu\nu}$ becomes
dynamical. The most general bimetric theory is given by \cite{HR5},
\bea
\label{bimetric}
S&=&M_g^2\int d^4x\sqrt{-\det g}\,R^{^{(g)}}+M_f^2\int d^4x\sqrt{-\det
  f}\,R^{^{(f)}} \nonumber \\ 
&&+2m^2 M_{\rm eff}^2 \int d^4x\sqrt{-\det g}\sum_{n=0}^{4} \beta_n\,
e_n(\sqrt{g^{-1} f}) \, . 
\eea
We have introduced two different Planck masses for the $f$ and $g$
sectors of the theory and defined an effective mass, 
\be
M^2_{\rm eff} =\Big(\frac{1}{M_g^2}+ \frac{1}{M_f^2}\Big)^{-1} \, .
\ee
The summation in the above action now runs from $n=0$ to $4$.  The
contribution of the new $n=4$ term is 
\be
\sqrt{-\det g}\,\beta_4 \, e_4(\sqrt{g^{-1} f}) = \beta_4 \sqrt{-\det f} \, .
\ee

Let us rewrite this action in terms of ADM variables, using the same
shift-like variables $n^i$ introduced in (\ref{NnD}), in place of
the shift $N^i$.  The Lagrangian becomes, 
\be
{\cal L} = M_g^2 \, \pi^{ij}\partial_t \gamma_{ij}+M_f^2 \,
p^{ij}\partial_t \tf_{ij}  - {\cal H}_0+N {\cal C}\, .
\label{Lbi}
\ee
Here ${\cal H}_0$ and ${\cal C}$ stand for 
\begin{align}
{\cal H}_0 &= -L^i\left(M_g^2 R_i^{^{(g)}}+M_f^2 R_i^{^{(f)}}\right)
-L\left(M_f^2 R^{0^{(f)}}+ M_g^2 n^iR_i^{^{(g)}} +2m^2 M^2_{\rm eff}
\sqrt{\det \gamma} \, U' \right) \, , 
\\[.1cm]
{\cal C} &= M_g^2 R^{0^{(g)}}+ M_g^2  R_i^{^{(g)}}D^i_{\,j}n^j +2m^2
M^2_{\rm eff} \sqrt{\det \gamma} \, V\,. \label{Cbi} 
\end{align}
In order to account for the new term, $\beta_4 \sqrt{-\det f}$, we
have introduced 
\be
\sqrt{\det \gamma} \, U' \equiv  \sqrt{\det \gamma} \, U+ \beta_4
\sqrt{\det \tf}  \, , 
\ee
where $U$ and $V$ in the above expressions are defined as before in
(\ref{U}) and (\ref{V}) respectively. 

The $p^{ij}$ are the momentum canonically conjugate to the dynamical variables $\tf_{ij}$.  The lapse $L$ and shift $L^i$ of $f_{\mu\nu}$ remain non-dynamical in the bimetric theory.  Thus the $\gamma_{ij}$, $\pi^{ij}$, $\tf_{ij}$ and $p^{ij}$ represent 24 potentially propagating phase-space degrees of freedom.   We argue now that the primary constraint ${\cal C}$ given above generates a secondary constraint, as in the massive gravity case, eliminating two phase-space degrees of freedom.  In addition, we will argue for the existence of eight additional constraints, related to the restored general coordinate invariance of the bimetric theory, that remove an additional eight phase-space degrees of freedom.  As a result, the bimetric theory has 14 phase-space degrees of freedom, or seven propagating modes, consistent with the seven modes one expects for a massive spin-2 field and a massless spin-2 field.

The bimetric constraint ${\cal C}$ in (\ref{Cbi}) is identical to that
defined above for massive gravity (\ref{C}) (up to irrelevant overall
factors), but now depends on the dynamical variables $\tf_{ij}$, both explicitly and  through the shift-like
variables $n^i$.  However, ${\cal C}$ is independent of $p^{ij}$, the momentum canonically to $\tf_{ij}$.  To see this, note that 
the $n^i$ equations of motion, determined from the Lagrangian (\ref{Lbi}), are
\begin{align}
&M_g^2 R_i^{^{(g)}}-2m^2M_{\rm eff}^2\sqrt{\det\gamma}\,
\frac{n^lf_{lj}}{\sqrt{x}}\bigg[\beta_1\,\delta^j_{i} 
+\beta_2\,\sqrt{x}\, (\delta^j_{i} D^m_{~m}-D^j_{~i})\nn \\
&\hspace{2.15cm}+\beta_3 \, \sqrt{x}^{\,2} \, \left(\tfrac{1}{2}
\delta^j_{~i}(D^m_{~m}D^n_{~n}\,-\,D^m_{~n}D^n_{~m})+D^j_{~m} D^m_{~i}\,-\,
D^j_{~i} D^m_{~m}\right) \bigg] =0\,.
\end{align}
These equations can be used to fix the $n^i$ in terms of the dynamical variables.  In addition to being independent of $N$, $L$ and $L^i$, these equations are also independent of $p^{ij}$.  Thus $n^i=n^i(\gamma_{ij}, \tf_{ij}, \pi^{ij})$.  As a result, the constraint ${\cal C}$ is independent of $p^{ij}$ as well.

So, when calculating the Poisson bracket $\pb{{\cal C}(x),{\cal C}(y)}$ for bimetric gravity, one can in fact consistently treat
the metric $f_{\mu\nu}$ as non-dynamical, since the bracket evaluated with respect to $(\tf_{ij},p^{ij})$ is zero. Thus the proof of existence of the secondary constraint in massive gravity holds for bimetric gravity as well. The primary
constraint ${\cal C}$ commutes with itself on the constraint surface.  Thus the consistency condition $d{\cal C}/dt = 0$ acts as a secondary constraint $\C$ on the dynamical variables, $\gamma_{ij}$, $\pi^{ij}$, $\tf_{ij}$ and $p^{ij}$.

In addition to these two constraints, we note that the Lagrangian
(\ref{Lbi}) is manifestly linear in the lapse $L$ and shift $L^i$.
Thus the variation of the action with respect to these variables
produces four additional constraints on the dynamical variables. These
four constraints, along with the four general coordinate invariances
of the bimetric theory and the primary and secondary constraints
${\cal C}$ and $\C$, reduce the initial 24 potentially dynamical phase-space
variables to 14, consistent with the seven modes of a massive spin-2
field coupled to a massless spin-2 field.

\vskip.5cm

\acknowledgments

We would like to thank F. Berkhahn, D. Blas, C. de Rham, K. Hinterbichler, N. Kaloper, F. K\"{u}hnel, J. Kluson, A. Schmidt-May, B. Sundborg, A. Tolley and M. Trodden for helpful discussions. We would also like to thank the organizers of the Workshop on Infrared Modifications of Gravity at ICTP, Trieste where the majority of this work was completed for creating a stimulating environment.  R.A.R.
is supported by NASA under contract NNX10AH14G.

\end{document}